\documentclass[preprint]{ptephy_v1}

\preprintnumber{KYUSHU-HET-194}

\usepackage{amsmath,amssymb}
\usepackage{mathtools}

\usepackage{graphicx}
\usepackage{booktabs}

\numberwithin{equation}{section}

\DeclareMathOperator{\tr}{tr}

\DeclareMathOperator{\sign}{sign}

\usepackage[%
pdftitle={Continuum limit in numerical simulations of the N=2 Landau--Ginzburg model},%
pdfsubject={},%
pdfauthor={Okuto Morikawa},%
pdfkeywords={Created on 26/5/2018; last updated on \today;%
}]{hyperref}

\title{Continuum limit in numerical simulations of the $\mathcal{N}=2$ Landau--Ginzburg model}
\author{Okuto Morikawa}
\affil{Department of Physics, Kyushu University,
744 Motooka, Nishi-ku, Fukuoka, 819-0395, Japan
\email{o-morikawa@phys.kyushu-u.ac.jp}
}

\begin{document}
\begin{abstract}
 The $\mathcal{N}=2$ Landau--Ginzburg description
 provides a strongly interacting Lagrangian realization
 of an~$\mathcal{N}=2$ superconformal field theory.
 It is conjectured that one such example is given by
 the two-dimensional~$\mathcal{N}=2$ Wess--Zumino model.
 Recently, the conjectured correspondence has been studied
 by using numerical techniques based on lattice field theory;
 the scaling dimension and the central charge have been directly measured.
 We study a single superfield with a cubic superpotential,
 and give an extrapolation method to the continuum limit.
 Then, on the basis of a supersymmetric-invariant numerical algorithm,
 we perform a precision measurement of the scaling dimension
 through a finite-size scaling analysis.
\end{abstract}
\subjectindex{B16, B24, B34, B38}
\maketitle

\tableofcontents
\section{Introduction}
A Lagrangian realization of a conformal field theory (CFT)
provides an important tool to clarify the conformal-invariant system.
As a famous example,
the Feigin--Fuks (integral) representation~\cite{Feigin:1981st,Feigin:1982tg}
gives a free-field Lagrangian on the curved spacetime.
Feigin and Fuks employed this
to explore the unitary representation of the Virasoro algebra,
and proved the Kac determinant formula in an elegant way.
Their technique has come in useful~\cite{Dotsenko:1984nm,Felder:1988zp}
for performing many computations explicitly and understanding the system intuitively.
Although the existence of such a Lagrangian is not always obvious,
one can extract more information from the Lagrangian
by using techniques based on quantum field theory.

A strongly interacting CFT Lagrangian
is realized by the Landau--Ginzburg (LG) model (or the LG description),
which is expected to become conformal invariant in extremely low-energy regions.
This realization is characterized
as a critical behavior under the renormalization group flow;
CFT would be a scale-invariant theory
on the non-trivial infrared (IR) fixed point under the flow.
Such critical phenomena are of great interest in a wide range of physics.
Originally, the idea of the LG description was introduced
as a phenomenological model to describe superconductivity~\cite{Ginzburg:1950sr};
in this context, the Lagrangian is replaced by the free energy.
To understand the critical behavior in an LG model
it is important to classify the critical exponent,
that is, the scaling of observables in the quantum field theory.

Let us consider one such example of LG models,
the two-dimensional ($2$D) massless~$\mathcal{N}=2$ Wess--Zumino (WZ)
model~\cite{Wess:1974tw} with a quasi-homogeneous superpotential.
From the dimensional reduction of the~$4$D~$\mathcal{N}=1$ WZ model,
the~$2$D~$\mathcal{N}=2$ WZ action with~$N_\Phi$ superfields is given by\footnote{%
Here, we consider the $2$D~$\mathcal{N}=(2,2)$ supersymmetry,
and not~$\mathcal{N}=(2,0)$.}
\begin{align}
 S &= \int d^2x \sum_{I=1}^{N_\Phi} \Biggl[
 4\partial A^*_I \Bar\partial A_I - F^*_I F_I
 - F^*_I \frac{\partial W(\{A\})^*}{\partial A_I^*}
 - F_I \frac{\partial W(\{A\})}{\partial A_I} \notag\\
 &\qquad\qquad\qquad
 + \left(\Bar\psi_{\Dot{1}}, \psi_2\right)_I \sum_{J=1}^{N_\Phi}
 \begin{pmatrix}
  2\delta_{IJ}\partial & \frac{\partial^2 W(\{A\})^*}{\partial A^*_I\partial A^*_J} \\ 
  \frac{\partial^2 W(\{A\})}{\partial A_I\partial A_J} & 2\delta_{IJ}\Bar\partial
 \end{pmatrix}
 \begin{pmatrix}
  \psi_1 \\ \Bar\psi_{\Dot{2}}
 \end{pmatrix}_J
 \Biggr],
 \label{eq:1.1}
\end{align}
where $A_I$ ($I=1$, \dots, $N_\Phi$) are complex scalars,
$\begin{pmatrix}\psi_1\\\Bar\psi_{\Dot{2}}\end{pmatrix}_I$ are $2$D Dirac fermions,
and $F_I$ are auxiliary fields;
we work in the Euclidean space, and
use the complex coordinate~$z=x_0+i x_1$ ($\Bar{z}=x_0-i x_1$)
and the corresponding derivative~$\partial=(\partial_0-i\partial_1)/2$
($\Bar\partial=(\partial_0+i\partial_1)/2$).
The model is believed to become an~$\mathcal{N}=2$ superconformal field theory (SCFT)
in the IR limit~\cite{DiVecchia:1985ief,DiVecchia:1986cdz,DiVecchia:1986fwg,%
Boucher:1986bh,Gepner:1986ip,Cappelli:1986hf,Cappelli:1986ed,Gepner:1986hr,%
Gepner:1987qi,Cappelli:1987xt,Kato:1987td,Gepner:1987vz}.
Much evidences of this conjectured WZ/SCFT correspondence has been given in~Refs.~\cite{%
Kastor:1988ef,Vafa:1988uu,Martinec:1988zu,Lerche:1989uy,Howe:1989qr,Cecotti:1989jc,%
Howe:1989az,Cecotti:1989gv,Cecotti:1990kz,Howe:1990cg,Witten:1993jg} and so on;
for example, Refs.~\cite{Kastor:1988ef,Howe:1989qr,Howe:1989az}
discuss the renormalization group flow for the~$N_\Phi=1$ WZ model
with the monomial superpotential, $W(\Phi)\propto\Phi^{n+1}$ ($n=2$, $3$, \dots),
which corresponds to the~$A_n$ minimal model of the~$\mathcal{N}=2$ SCFT.
See~Refs.~\cite{Polchinski:1998rr,Hori:2003ic,Tachikawa:2018sae} for reviews.
However, we have no complete proof of the conjectured LG correspondence to SCFT.
This is because
the $2$D~$\mathcal{N}=2$ WZ model is strongly coupled in low-energy regions,
and perturbation theory possesses IR divergences.
It is difficult to directly observe the critical behavior in the WZ model.

Recently, the conjectured WZ/SCFT correspondence has been non-perturbatively studied
by using numerical techniques based on lattice field theory.
In the case of a single superfield with cubic and quartic superpotentials,
which corresponds to the~$A_2$ and~$A_3$ minimal models, respectively,
the authors of~Refs.~\cite{Kawai:2010yj,Kamata:2011fr,Morikawa:2018ops}
numerically measured the scaling dimension~$h+\Bar{h}$ of the primary fields
(see~Table~\ref{tab:scaling_dim}).
\begin{table}[t]
 \centering
 \caption{Scaling dimension~$1-h-\Bar{h}$ measured in preceding studies.}
 \label{tab:scaling_dim}
 \begin{tabular}{rlll}\toprule
  & Reference & $1-h-\Bar{h}$ & Expected value \\
  \midrule
  $A_2$ & Kawai--Kikukawa~\cite{Kawai:2010yj} & $0.660(11)$ & $2/3=0.666\dots$ \\
        & Kamata--Suzuki~\cite{Kamata:2011fr} & $0.616(25)(13)$ & \\
        & Morikawa--Suzuki~\cite{Morikawa:2018ops} & $0.682(10)(7)$ & \\
  \midrule
  $A_3$ & Morikawa--Suzuki~\cite{Morikawa:2018ops} & $0.747(11)(12)$ & $0.75$ \\
  \bottomrule
 \end{tabular}
\end{table}
The first remarkable study~\cite{Kawai:2010yj}
is based on the lattice formulation by Kikukawa and Nakayama~\cite{Kikukawa:2002as},
which preserves one nilpotent supersymmetry (SUSY) exactly;\footnote{%
In the continuum limit,
the full SUSY in the formulation~\cite{Kikukawa:2002as} is automatically restored
to all orders of perturbation theory~\cite{Giedt:2004qs,Kadoh:2010ca}.
Reference~\cite{Kadoh:2016eju} is a review of SUSY on the lattice,
which refers to lattice formulations of the~$2$D~$\mathcal{N}=2$ WZ model.}
the others are
on the \textit{SUSY-preserving} momentum-cutoff regularization
by Kadoh and Suzuki~\cite{Kadoh:2009sp}.
Both non-perturbative formulations make essential use of
the existence of the Nicolai or Nicolai--Parisi--Sourlas
mapping~\cite{Nicolai:1979nr,Nicolai:1980jc,Parisi:1982ud,Cecotti:1983up}.
In particular, by applying the latter formulation
to the WZ model with multiple superfields,
the central charge in ADE-type minimal models~\cite{Vafa:1988uu} had also been measured
quite straightforwardly~\cite{Kamata:2011fr,Morikawa:2018ops,Morikawa:2018zys}.
One can observe good agreement of
the scaling dimension~$h+\Bar{h}$ in~Table~\ref{tab:scaling_dim}
and the central charge~\cite{Kamata:2011fr,Morikawa:2018ops,Morikawa:2018zys}
with those of the expected minimal models.
These studies achieved a triumph for lattice field theory,
and enable us to study more general~$\mathcal{N}=2$ SCFTs.

Although the corresponding SCFT is defined as the continuum theory with infinite volume,
the above results are not extrapolated to the thermodynamic and continuum limits.
Moreover, it was noted~\cite{Morikawa:2018ops}
that the computation of~$h+\Bar{h}$ in~Ref.~\cite{Kamata:2011fr} is
quite sensitive to a UV ambiguity
because of the locality breaking in the Kadoh--Suzuki formulation with a finite cutoff.
To justify numerical studies based on the formulation,
such a UV ambiguity should disappear in the infinite-volume and continuum limits.
It is important and helpful
to analyze the limits and precisely determine the scaling dimension.

In this paper we study a single superfield with the cubic superpotential
on the basis of the SUSY-invariant formulation,
which is believed to correspond to the~$A_2$ minimal model.
The finite-size scaling analysis in~Refs.~\cite{Kamata:2011fr,Kawai:2010yj}
is developed into an analysis method with continuum-limit extrapolation.
The extrapolation also carries out the thermodynamic limit.
Then, we numerically simulate the IR behavior of a scalar correlator,
extrapolate it to the continuum limit,
and perform a precision measurement of the scaling dimension;
we have the scaling dimension
\begin{align}
 1 - h - \Bar{h} &= 0.6699(77)(87) .
 \label{eq:1.2}
\end{align}
This more reliable result is rather consistent with the conjectured~$A_2$-type correspondence.
Our computation would support the restoration of the locality in the continuum limit.
In this regard, the theoretical background of the formulation is still not clear,
so the restoration of the locality should be observed more carefully.
One can apply our extrapolation method to other non-perturbative formulations.
We hope that the numerical approaches, when further developed,
will be useful to investigate a superstring theory through the LG/Calabi--Yau
correspondence~\cite{Martinec:1988zu,Cecotti:1990wz,Greene:1988ut,Witten:1993yc}.

\section{SUSY-preserving formulation}
We consider the~$A$-type theory, that is,
the~$N_\Phi=1$ WZ model of~Eq.~\eqref{eq:1.1} with the superpotential
\begin{align}
 W(\Phi) = \frac{\lambda}{n+1} \Phi^{n+1},
 \label{eq:2.1}
\end{align}
where $n$ is a positive integer, $\lambda$ is a dimensionful coupling,
and we have omitted the index~$I$ from the field variable;
the theory is conjectured to correspond to the~$A_n$ minimal model.
Let us suppose that the system is defined
in a $2$D Euclidean box of physical size~$L_0 \times L_1$.
Then, the Fourier transformation of each field~$\varphi(x)$ is defined by
\begin{align}
 \varphi(x) &= \frac{1}{L_0L_1} \sum_p e^{ip\cdot x} \varphi(p), &
 \varphi(p) &= \int d^2x\, e^{-ip\cdot x} \varphi(x) .
 \label{eq:2.2}
\end{align}
Here, the momentum~$p$ is discretized as
\begin{align}
 p_\mu = \frac{2\pi}{L_\mu} n_\mu
 \qquad
 (n_\mu = 0,\pm1,\pm2,\dots) ,
 \label{eq:2.3}
\end{align}
where the Greek index~$\mu$ runs over~$0$ and~$1$,
and repeated indices are not summed over.
Integrating over the auxiliary field~$F$,
we obtain the action in terms of the Fourier modes of the physical component fields,
\begin{align}
 S = S_B
 + \frac{1}{L_0L_1} \sum_p \left(\Bar\psi_{\Dot{1}},\psi_2\right)(-p)
 \begin{pmatrix}
  2ip_z & W''(A)^* * \\ W''(A) * & 2ip_{\Bar{z}}
 \end{pmatrix}
 \begin{pmatrix}
  \psi_1 \\ \Bar\psi_{\Dot{2}}
 \end{pmatrix} (p),
 \label{eq:2.4}
\end{align}
where $p_z=(p_0-ip_1)/2$ ($p_{\Bar{z}}=(p_0+ip_1)/2$),
the symbol~$*$ denotes the convolution
\begin{align}
 (\varphi_1*\varphi_2)(p)
 \equiv \frac{1}{L_0L_1} \sum_q \varphi_1(q) \varphi_2(p-q),
 \label{eq:2.5}
\end{align}
and the boson part of the action, $S_B$, is given by
\begin{align}
 S_B &\equiv \frac{1}{L_0L_1} \sum_p N^*(-p) N(p),&
 N(p) &\equiv 2ip_z A(p) + W'(A)^* (p).
 \label{eq:2.6}
\end{align}
The field products in~$W'(A)$ and~$W''(A)$ are understood as the convolution.
The new variable~$N(p)$ in~Eq.~\eqref{eq:2.6} specifies the so-called Nicolai
mapping~\cite{Nicolai:1979nr,Nicolai:1980jc,Parisi:1982ud,Cecotti:1983up};
the change of variables from~$A$ to~$N$ simplifies the path-integral weight drastically,
as we will see soon.

In what follows,
we employ a momentum-cutoff regularization given in~Ref.~\cite{Kadoh:2009sp}.
In the formulation, a momentum cutoff~$\Lambda$ is introduced as
\begin{align}
 |p_\mu| \leq \Lambda
 \qquad \text{for $\mu=0$ and $1$}.
 \label{eq:2.7}
\end{align}
Then, we also define a ``lattice spacing''~$a$ by
\begin{align}
 \Lambda \equiv \frac{\pi}{a} ,
 \label{eq:2.8}
\end{align}
and all dimensionful quantities are measured in units of~$a$.
Although an underlying lattice space is not always assumed~\cite{Kamata:2011fr},
we will use this parameter to take the ``continuum limit''~$a\to0$,
which implies that we remove the UV cutoff as $\Lambda\to\infty$.
The partition function is then given by
\begin{align}
 \mathcal{Z}
 &= \int\prod_{|p_\mu|\leq\frac{\pi}{a}}
 \left[dA(p)dA^*(p)
 \prod_{\alpha=1}^{2}d\psi_\alpha(p)
 \prod_{\Dot\alpha=\Dot{1}}^{\Dot{2}}d\Bar\psi_{\Dot\alpha}(p)\right]
 e^{-S}
 \notag\\
 &= \int\prod_{|p_\mu|\leq\frac{\pi}{a}} \left[dN(p)dN^*(p)\right]
 e^{-S_B} \sum_i
 \left.\sign \det \frac{\partial(N,N^*)}{\partial(A,A^*)}\right|_{A=A_i,A^*=A_i^*},
 \label{eq:2.9}
\end{align}
where $A_i$ ($i=1$, $2$, \dots) are solutions of the equation
\begin{align}
 2ip_z A(p) + W'(A)^*(p) - N(p) = 0,
 \label{eq:2.10}
\end{align}
and $A_i^*$ are their complex conjugates.
In the second line of~Eq.~\eqref{eq:2.9},
we have used the Nicolai mapping in~Eq.~\eqref{eq:2.6}
and integrated over the fermion fields;
note that the fermion determinant coincides
with the Jacobian associated with the Nicolai mapping, up to the sign:
\begin{align}
 \det
 \begin{pmatrix}
  2ip_z & W''(A)^* * \\ W''(A) * & 2ip_{\Bar{z}}
 \end{pmatrix}
 = \det \frac{\partial(N,N^*)}{\partial(A,A^*)} .
 \label{eq:2.11}
\end{align}
The simulation algorithm is summarized
in~Refs.~\cite{Kamata:2011fr,Morikawa:2018ops,Morikawa:2018zys}.

This regularized system, Eq.~\eqref{eq:2.9}, possesses some remarkable features:
\begin{enumerate}
 \item
 This regularization exactly preserves
 SUSY, the translational invariance, and the~$U(1)$ symmetry.
 Thus, we can quite straightforwardly construct the appropriate expression for
 the supercurrent, the energy-momentum tensor, and the~$U(1)$ current such that
 they form the~$\mathcal{N}=2$ superconformal multiplet~\cite{Morikawa:2018ops}.
 This fact enables us to numerically compute such Noether currents
 directly and easily~\cite{Kamata:2011fr,Morikawa:2018ops,Morikawa:2018zys}.\footnote{%
 See~Refs.~\cite{Caracciolo:1989pt,Caracciolo:1991cp}
 for a general construction of the energy-momentum tensor in lattice field theory.
 Recently, a regularization-independent construction of such Noether currents
 has been developed in terms of the gradient
 flow~\cite{Narayanan:2006rf,Luscher:2009eq,Luscher:2010iy,Luscher:2011bx};
 see also~Ref.~\cite{Suzuki:2016ytc} for a review.}
 \item
 The path-integral weight~$\exp(-S_B)$ is a Gaussian function of~$N(p)$.
 Thus we can obtain configurations of~$N(p)$
 by generating Gaussian random numbers for each~$p_\mu$.
 This algorithm is completely free from any undesired autocorrelation
 and the critical slowing down.
 \item
 The normalized partition function,
 \begin{align}
  \Delta =
  \left\langle \sum_i
  \left.\sign \det \frac{\partial(N,N^*)}{\partial(A,A^*)}\right|_{A=A_i,A^*=A_i^*}
  \right\rangle,
  \label{eq:2.12}
 \end{align}
 can be computed numerically, which gives
 the Witten index, $\tr(-1)^F$~\cite{Witten:1982df,Cecotti:1981fu}.
 When the superpotential is a polynomial of degree~$n$, e.g. $W(A)\propto A^{n+1}$,
 we should have~$\Delta = n$.
\end{enumerate}

Unfortunately, there are some difficulties for the algorithm;
see, e.g., Ref.~\cite{Morikawa:2018ops}.
In particular, the momentum cutoff breaks the locality of the theory.
When the numbers $L_\mu/a$ are taken as odd integers,
this formulation is nothing but the dimensional reduction
of the lattice formulation of the $4$D WZ model~\cite{Bartels:1983wm}
based on the SLAC derivative~\cite{Drell:1976bq,Drell:1976mj};
this is plagued by the pathology that the locality is not automatically restored
in the continuum limit~\cite{Dondi:1976tx,Karsten:1979wh,Kato:2008sp,Bergner:2009vg}.
On the other hand, for the \textit{massive} $2$D~$\mathcal{N}=2$ WZ model,
one can argue the restoration of it as~$a\to0$
within perturbation theory~\cite{Kadoh:2009sp}.
For the \textit{massless} case, since perturbation theory possesses IR divergences,
it is not clear whether its restoration is automatically accomplished.
Nevertheless, the numerical results in the preceding studies and ours below
suggest the validity of the approach.

\section{Numerical setup}\label{sec:setup}
We summarize the numerical setup that we will use in this paper.
Our setup is based on the simulation setup in~Ref.~\cite{Morikawa:2018ops}.
We consider the~$2$D~$\mathcal{N}=2$ WZ model
with the superpotential of~Eq.~\eqref{eq:2.1} of degree~$2$,
\begin{align}
 W(\Phi) &= \frac{\lambda}{3} \Phi^3,
 \label{eq:3.1}
\end{align}
which corresponds to the~$A_2$ minimal model.
Here, the coupling constant~$\lambda$ is a dimensionful parameter
and characterizes the mass scale in this theory.
For simplicity, the system is supposed to be defined in the physical box~$L \times L$,
where $L/a$ is taken as even integers in the interval~$[10,52]$.

To numerically compute observables, e.g. ~Eq.~\eqref{eq:2.12},
we first generate Gaussian random numbers~$N(p)$ for each~$p_\mu$.
Then we solve the multi-variable algebraic equation in~Eq.~\eqref{eq:2.10}
with respect to~$A(p)$;
we should ideally find all the solutions~$A_i(p)$ ($i=1$, $2$, $\dots$) numerically.
To do this, we employ the Newton--Raphson method
and set the convergence threshold as
\begin{align}
 \sqrt{\frac{\sum_p|2ip_z A(p)+W'(A)^*(p)-N(p)|^2}{\sum_q|N(q)|^2}} <
 \begin{cases}
  10^{-14} & \text{for $L<52a$} \\
  10^{-13} & \text{for $L=52a$} .
 \end{cases}
 \label{eq:3.2}
\end{align}
In the case of~$L=52a$, which is the most numerically demanding one in this paper,
the threshold is less accurate
(and also the number of obtained configurations is not relatively high).
For a configuration~$N(p)$, we randomly generate initial trial configurations of~$A(p)$
by Gaussian random numbers with unit variance,
so that we obtain $200$ solutions for~$A$, allowing repetition of identical solutions,
with~$L<52a$ and $120$ solutions with~$L=52a$.
Two solutions~$A_1$ and~$A_2$ are regarded as identical if
\begin{align}
 \sqrt{\frac{\sum_p|A_1(p)-A_2(p)|^2}{\sum_q|A_1(q)|^2}} <
 \begin{cases}
  10^{-11} & \text{for $L<52a$} \\
  10^{-10} & \text{for $L=52a$} .
 \end{cases}
 \label{eq:3.3}
\end{align}

Finally, we tabulate the classification of the configurations obtained
in~Table~\ref{tab:classification},
where the coupling~$a\lambda$ has already been tuned
in accordance with an argument given in the next section.
In~Table~\ref{tab:quality},
we list the numerical results of the Witten index in~Eq.~\eqref{eq:2.12}, $\Delta=2$,
and the one-point SUSY Ward--Takahashi identity~\cite{Catterall:2001fr}
(see also~Ref.~\cite{Morikawa:2018ops})
\begin{align}
 \delta \equiv \frac{\left\langle S_B \right\rangle}{(L+1)^2} - 1 = 0.
 \label{eq:3.4}
\end{align}
Whether~$\Delta$ and~$\delta$ are numerically reproduced
indicates the quality of our configurations.
\begin{table}[t]
 \centering
 \caption{Classification of the configurations obtained for the~$A_2$-type theory.
 $\mathcal{N}_\text{conf}$ denotes the total number of configurations for each setup.
 In the upper half of the table,
 the number of configurations for~$L$ is shown;
 in the lower half, that for~$L'=2L$ is shown.
 The symbol~$(n,m)$ implies that,
 for a configuration~$N(p)$, we find $(n+m)$ solutions, $A_i(p)$ ($i=1$, \dots, $n+m$);
 the~$n$ solutions take~$\det\sign\frac{\partial(N,N^*)}{\partial(A,A^*)}=+1$
 and the~$m$ solutions take~$-1$.
 }
 \label{tab:classification}
 \begin{tabular}{rlrrrrrrrrr}\toprule
  $L/a$ & $a\lambda$ & $\mathcal{N}_\text{conf}$ & $(2,0)$ & $(3,1)$ & $(4,2)$
  & $(1,0)$ & $(2,1)$ & $(3,2)$ & $(3,0)$ & $(4,1)$
  \\\midrule
  10 & 0.1780 & 7680 & 7680 & 0   & 0 & 0 & 0 & 0 & 0 & 0 \\
  12 & 0.2135 & 5120 & 5119 & 1   & 0 & 0 & 0 & 0 & 0 & 0 \\
  14 & 0.2538 & 5120 & 5119 & 1   & 0 & 0 & 0 & 0 & 0 & 0 \\
  16 & 0.3000 & 5120 & 5112 & 8   & 0 & 0 & 0 & 0 & 0 & 0 \\
  18 & 0.3420 & 5120 & 5093 & 27  & 0 & 0 & 0 & 0 & 0 & 0 \\
  20 & 0.3888 & 5120 & 5070 & 50  & 0 & 0 & 0 & 0 & 0 & 0 \\
  22 & 0.4500 & 5120 & 5023 & 97  & 0 & 0 & 0 & 0 & 0 & 0 \\
  24 & 0.5100 & 5120 & 4961 & 156 & 3 & 0 & 0 & 0 & 0 & 0 \\
  26 & 0.5705 & 5120 & 4909 & 204 & 6 & 0 & 0 & 0 & 1 & 0 \\
  \midrule
  20 & 0.1780 & 5120 & 5117 & 3   & 0 & 0 & 0 & 0 & 0 & 0 \\
  24 & 0.2135 & 5120 & 5104 & 16  & 0 & 0 & 0 & 0 & 0 & 0 \\
  28 & 0.2538 & 5120 & 5075 & 44  & 1 & 0 & 0 & 0 & 0 & 0 \\
  32 & 0.3000 & 4320 & 4236 & 83  & 1 & 0 & 0 & 0 & 0 & 0 \\
  36 & 0.3420 & 2592 & 2514 & 77  & 1 & 0 & 0 & 0 & 0 & 0 \\
  40 & 0.3888 & 2592 & 2472 & 118 & 0 & 0 & 1 & 1 & 0 & 0 \\
  44 & 0.4500 & 2592 & 2458 & 131 & 2 & 0 & 0 & 0 & 0 & 1 \\
  48 & 0.5100 & 2592 & 2433 & 157 & 2 & 0 & 0 & 0 & 0 & 0 \\
  52 & 0.5705 & 1512 & 1392 & 107 & 4 & 1 & 1 & 1 & 6 & 0 \\
  \bottomrule
 \end{tabular}
\end{table}
\begin{table}[t]
 \centering
 \caption{Quality of the configurations obtained for the~$A_2$-type theory.
 The Witten index of~Eq.~$\Delta$~\eqref{eq:2.12} and the one-point
 function of~Eq.~$\delta$~\eqref{eq:3.4} are numerically computed for~$L$ and~$L'=2L$;
 $\Delta$ should be identical to~$2$, and~$\delta$ should identically vanish.
 For~$L'/a=52$,
 the quality of the configurations obtained is poorer due to the computational cost.}
 \label{tab:quality}
 \begin{tabular}{rrlllll}\toprule
  $L/a$ & $L'/a$ & $a\lambda$ & $\Delta(L)$ & $\Delta(L')$ & $\delta(L)$ & $\delta(L')$
  \\\midrule
  10 & 20 & 0.1780 & 2         & 2         & $-$0.00099(104) & $-$0.00005(67)\\
  12 & 24 & 0.2135 & 2         & 2         & $-$0.00063(107) & $+$0.00046(56)\\
  14 & 28 & 0.2538 & 2         & 2         & $-$0.00019(94)  & $-$0.00030(48)\\
  16 & 32 & 0.3000 & 2         & 2         & $-$0.00024(81)  & $-$0.00004(46)\\
  18 & 36 & 0.3420 & 2         & 2         & $-$0.00109(74)  & $+$0.00020(52)\\
  20 & 40 & 0.3888 & 2         & 1.9992(5) & $-$0.00078(67)  & $+$0.00053(55)\\
  22 & 44 & 0.4500 & 2         & 2.0004(4) & $-$0.00005(62)  & $+$0.00031(48)\\
  24 & 48 & 0.5100 & 2         & 2         & $+$0.00041(56)  & $+$0.00000(41)\\
  26 & 52 & 0.5705 & 2.0002(2) & 2.002(2)  & $-$0.00058(52)  & $+$0.00073(110)\\
  \bottomrule
 \end{tabular}
\end{table}
\clearpage

\section{Scaling dimension}
\subsection{Susceptibility of the scalar field}
To numerically determine the scaling dimension,
we first explain the finite-size scaling analysis
in~Refs.~\cite{Kawai:2010yj,Kamata:2011fr},
which is compatible with the continuum limit as we will develop later.
Let us consider the susceptibility of the scalar field~$A$,
defined by~\cite{Kawai:2010yj}
\begin{align}
 \chi(L_\mu)
 = \frac{1}{a^2} \int_{L_0 L_1} d^2 x\, \left\langle A(x) A^{*}(0) \right\rangle
 = \frac{1}{a^2L_0L_1} \left\langle |A(p=0)|^2 \right\rangle.
 \label{eq:4.1}
\end{align}
In the IR limit, the scalar field is expected to behave as a chiral primary field
with the conformal dimensions~$(h, \Bar{h})$;
the two-point function of~$A$ behaves as
\begin{align}
 \left\langle A(x) A^{*}(0) \right\rangle = \frac{1}{z^{2h} \Bar{z}^{2\Bar{h}}},
 \label{eq:4.2}
\end{align}
for large~$|x| = \sqrt{x^2}$.
Note that $h+\Bar{h}$ is called the scaling dimension, and $h-\Bar{h}$ is the spin.
Now suppose that the field~$A$ is spinless, $h=\Bar{h}$.
Then, we observe the finite-volume scaling of the scalar susceptibility
for large~$L_\mu$, as
\begin{align}
 \chi \propto (L_0 L_1)^{1 - h - \Bar{h}}.
 \label{eq:4.3}
\end{align}
Numerically simulating the scalar correlator for some different volumes
but the same value of the coupling, one can read the exponent, $1-h-\Bar{h}$,
from the slope of~$\ln\chi(L_\mu)$ as a linear function of~$\ln(L_0L_1)$.
In what follows, for simplicity,
we take into account the case of the physical box size~$L=L_0=L_1$.

\subsection{Continuum limit of the susceptibility}\label{sec:cont-lim}
As already announced, we consider the thermodynamic and continuum limits, $a/L\to0$.
No extrapolation has been done in the preceding numerical studies.
In~Refs.~\cite{Kawai:2010yj,Kamata:2011fr,Morikawa:2018ops,Morikawa:2018zys},
the grid size~$L/a$ is expected to be taken as sufficiently large values,
while the coupling~$\lambda$ in the superpotential in~Eq.~\eqref{eq:2.1} is fixed
by~$a\lambda = 0.3$;
good agreement of the scaling dimension
with those of the~$A_2$ and~$A_3$ minimal models was observed
(Table~\ref{tab:scaling_dim}).
Unlike in the case of QCD, however,
the present model does not possess any dynamical scale,
so the ``sufficiently small'' scale of~$a$ is not obvious.
In fact, we will find that
the susceptibility, $\chi(L)$, takes a slow approach to~$a/L=0$.
To obtain precise and reliable results,
we should extend the above finite-size scaling analysis
in order to treat the thermodynamic and continuum limits.

We have also recognized the pathology of the locality in the lattice formulation
that is based on the SLAC derivative;
the computation of~$\ln\chi(L)$ with finite~$L/a$ is
quite sensitive to this problem~\cite{Kamata:2011fr,Morikawa:2018ops}
(see also~Sect.~\ref{sec:uv-subt}).
A proposal given in~Ref.~\cite{Morikawa:2018ops}
is to directly study the correlation function in the momentum space,
$\langle A(p)A^*(-p) \rangle$.
Although the measured scaling dimension with the fixed coupling
tends to approach expected values as the grid size~$L/a$ increases,
the approach to the~$L/a\to\infty$ limit appears not quite
smooth~\cite{Morikawa:2018ops}.\footnote{%
The central charge, which can be measured by computing
the energy-momentum tensor correlator~$\langle T_{zz}(p)T_{zz}(-p) \rangle$,
appears to possess a higher convergence speed than the scaling dimension,
though the approach to~$L/a\to\infty$ is also
not quite smooth~\cite{Morikawa:2018ops,Morikawa:2018zys}.}
We would need a more systematic method for the infinite-volume and continuum limits,
while the locality should be restored in the limits. 

Our strategy for the continuum limit
is very similar to that in~Ref.~\cite{Luscher:1991wu}.
We regard~$\ln\chi(L)$ as the same kind of
running coupling~$\Bar{g}^2(L)$ defined on a lattice.
To take the continuum limit, various sizes of the lattice spacing
$\{a_i\}$ ($i=1$, $2$, \dots) are required;
we first prepare various momentum-grid sizes~$\{L/a_i\}$,
while the lattice parameter~$a_i\lambda$ is tuned so that
$\ln\chi(L)$ (or~$\Bar{g}^2(L)$) is kept fixed; we denote $u=\ln\chi(L)$.
A system with a different grid size~$L'/a'\neq L/a_i$
and the same parameter~$a'\lambda'=a_i\lambda$
possesses the physical box size~$L'\times L'$ with~$a'=a_i$.
Then, we compute~$\ln\chi(L')$ ($\Bar{g}^2(L')$) for~$L'/a_i$ and~$a_i\lambda$;
we observe the $a$-dependence of~$\ln\chi(L')|_a$ ($\Bar{g}^2(L')|_a$),
and attempt to extrapolate this in the continuum limit, $\lim_{a\to0} \ln\chi(L')|_a$.

To be more specific, we introduce the scaling function~$\Sigma$ as
\begin{align}
 \Sigma(s, u, a/L) = \ln \chi(s L)|_a.
 \label{eq:4.4}
\end{align}
The statistical error of~$\Sigma$ would be given by
the square root of the sum of squared errors of~$\ln\chi(L)$ and~$\ln\chi(s L)$,
owing to the long-distance behavior in~Eq.~\eqref{eq:4.3}.
As a consequence of the continuum limit with a to-be-determined fit function,
we can obtain the scaling dimension
\begin{align}
 1 - h - \Bar{h}
 &= \frac{1}{\ln s^2} \left[ \lim_{a\to0} \Sigma(s, u, a/L) - u \right] .
 \label{eq:4.5}
\end{align}
The cutoff dependence will be determined from numerical results.
Note that the unique mass scale~$\lambda$ in this model should be sufficiently larger
than~$1/L$ to study the conformal behavior~\cite{Kawai:2010yj},
hence~$\lambda L\to\infty$ as the continuum limit.
This indicates that the extrapolation carries out the thermodynamic limit
at the same time.
We can apply our extrapolation method to the continuum limit
to other non-perturbative formulations,
for example the lattice formulation in~Ref.~\cite{Kawai:2010yj}.

\subsection{Numerical measurement of the scaling dimension}
In this subsection, we perform precision measurement of the scaling dimension
for the~$A_2$-type theory with the cubic superpotential~$\Phi^3$
by using the above continuum-limit analysis.
In~Sect.~\ref{sec:setup} we had already summarized our parameter set
and the classification of the obtained configurations.

We tabulate the numerical results of the scalar susceptibility
with the various box sizes of~$L$ and~$L'=2L$ in~Table~\ref{tab:chi}.
The third column is devoted to the tuned values of the coupling, $a\lambda$,
so that~$\ln\chi(L)$ in the fourth column is kept almost fixed.
The results of~$\Sigma(u,a/L)$ are shown in the last column,
where we have omitted the first argument~$s=2$ of~$\Sigma(s,u,a/L)$,
while we set~$u=\ln\chi(L)$ as~$3.9175$.
The error of~$\Sigma(u,a/L)$ is given by
the square root of the sum of the squared errors of~$\ln\chi(L)$ and~$\ln\chi(L')$.
\begin{table}[t]
 \centering
 \caption{Scalar susceptibility, $u=3.9175$.}
 \label{tab:chi}
 \begin{tabular}{rrllll}\toprule
  $L/a$ & $L'/a$ & $a\lambda$ & $\ln\chi(L)$ & $\ln\chi(L')$ & $\Sigma(u,a/L)$
  \\\midrule
  10 & 20 & 0.1780 & 3.9174(59) & 4.6338(72)  & 4.6338(93)  \\
  12 & 24 & 0.2135 & 3.9175(73) & 4.6642(69)  & 4.6642(100) \\
  14 & 28 & 0.2538 & 3.9193(70) & 4.6844(66)  & 4.6844(97)  \\
  16 & 32 & 0.3000 & 3.9171(69) & 4.6913(68)  & 4.6913(97)  \\
  18 & 36 & 0.3420 & 3.9166(68) & 4.7223(83)  & 4.7223(107) \\
  20 & 40 & 0.3888 & 3.9215(65) & 4.7251(81)  & 4.7251(104) \\
  22 & 44 & 0.4500 & 3.9162(62) & 4.7400(76)  & 4.7400(97)  \\
  24 & 48 & 0.5100 & 3.9186(60) & 4.7610(70)  & 4.7610(93)  \\
  26 & 52 & 0.5705 & 3.9175(56) & 4.7823(91)  & 4.7823(107) \\
  \bottomrule
 \end{tabular}
\end{table}

In~Ref.~\cite{Kamata:2011fr} the scaling dimension was obtained
from the slope of the susceptibility in the formulation by using data
for~$24\leq L/a \leq 36$ or~$26\leq L/a \leq 36$ with a fixed coupling;
we have a similar slope of~$\ln\chi$ for~$(L/a,L'/a)=(24,48)$,
though we have used different values of~$a\lambda$ (see~Table~\ref{tab:h_finite}).
We will find a significant difference
between such numerical results at a finite cutoff and our result below at~$a/L=0$.
\begin{table}[t]
 \centering
 \caption{Scaling dimension measured at finite volumes.
 The results in the last two rows are obtained
 by reading the slope of~$\ln\chi$ for~$(L/a,L'/a)=(24,48)$ or~$(L/a,L'/a)=(26,52)$
 in~Table~\ref{tab:chi}.}
 \label{tab:h_finite}
 \begin{tabular}{llll}\toprule
  & Fit range of~$L$ & $a\lambda$ & $1-h-\Bar{h}$ \\
  \midrule
  Kamata--Suzuki~\cite{Kamata:2011fr}
  & From~24 to~36 & 0.3000 & 0.603(19)\\
  & From~26 to~36 & 0.3000 & 0.609(25)\\
  \midrule
  & From~24 to~48 & 0.5100 & 0.6076(66)\\
  & From~26 to~52 & 0.5705 & 0.6238(77) \\
  \bottomrule
 \end{tabular}
\end{table}

\begin{figure}[t]
 \begin{center}
  \includegraphics[width=0.8\columnwidth]{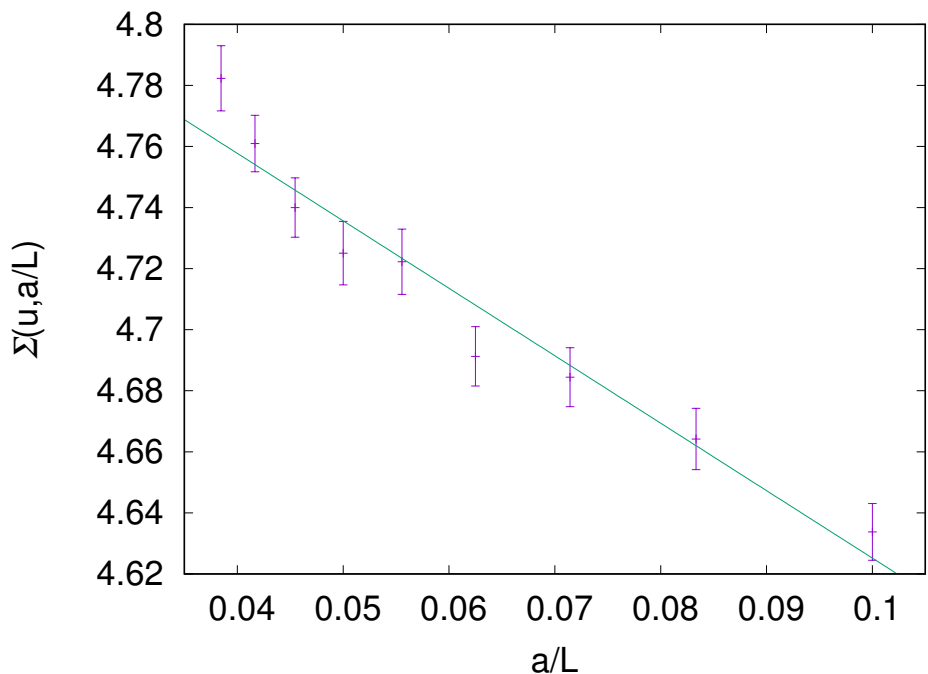}
  \caption{$\Sigma(u,a/L)$-$(a/L)$ plot with $u=3.9175$.
   The fitting line of~Eq.~\eqref{eq:4.6} is also depicted.}
  \label{fig:chi}
 \end{center}
\end{figure}

Now we have enough data to clarify the $(a/L)$-dependence of~$\Sigma(u,a/L)$.
Figure~\ref{fig:chi} shows~$\Sigma(u,a/L)$ as a function of~$a/L$
given in~Table~\ref{tab:chi}.
From the plot, we simply apply a linear function of~$a/L$
in order to take the continuum limit; then we have
\begin{align}
 \Sigma(3.9175,a/L) &= -0.0850(64) \times \frac{26a}{L} + 4.8461(107) ,
 \label{eq:4.6}
\end{align}
with~$\chi^2/\text{d.o.f.} = 1.417$.
From~Eq.~\eqref{eq:4.5}, the scaling dimension is given by
\begin{align}
 1-h-\Bar{h} &= 0.6699(77) .
 \label{eq:4.7}
\end{align}
This result is consistent with the expected exact value~$1-h-\Bar{h}=2/3=0.6666\dots$
within the statistical error.

Because the quality of configurations with $L/a=52$ is poorer
due to the computational cost (see~Sect.~\ref{sec:setup}),
the computation of~$\ln\chi$ could be less accurate.
In fact, the above result in Fig.~\ref{fig:chi} implies that
there is a discrepancy between the central values of~$\ln\chi(L)$
and the fit function at~$L/a=52$.
To make sure that this discrepancy comes from statistical fluctuations,
we show the behavior of~$\ln\chi(L)$ for~$L/a=52$
when the number of configurations varies in~Table~\ref{tab:stat_chi};
the deviation of the central values decreases.
\begin{table}[t]
 \centering
 \caption{$\ln\chi(L')$ with~$u=3.9175$ and $L/a=52$
 when the number of configurations, $\mathcal{N}_\text{conf}$, varies.}
 \label{tab:stat_chi}
 \begin{tabular}{rl}\toprule
  $\mathcal{N}_{\text{conf}}$ & $\ln\chi(L')$
  \\\midrule
  1512 & 4.7823(91) \\
  756  & 4.7950(133)\\
  378  & 4.8087(193)\\
  \bottomrule
 \end{tabular}
\end{table}

To estimate the systematic error, we may omit the configurations for~$L/a=52$; that is,
\begin{align}
 \left.\Sigma(3.9175,a/L)\right|_{L/a<52}
 &= -0.0791(69) \times \frac{26a}{L} + 4.8341(120) ,
 \label{eq:4.8}
\end{align}
with~$\chi^2/\text{d.o.f.} = 0.807$; we obtain
\begin{align}
 1-h-\Bar{h}=0.6612(86) .
 \label{eq:4.9}
\end{align}
The main result of the scaling dimension in this paper is given by
\begin{align}
 1 - h - \Bar{h} = 0.6699(77)(87) .
 \label{eq:4.10}
\end{align}
Here, the number in the second parentheses indicates the systematic error
defined by the deviation between the central values
of~Eq.~\eqref{eq:4.7} and~Eq.~\eqref{eq:4.9}.

\subsection{Discussion on the fit function}\label{sec:uv-subt}
We found that a linear fit of~$\Sigma(s,u,a/L)$ with respect to~$a/L$ would be good
within the numerical error.
To convince ourselves of this fact,
let us introduce a slightly modified extrapolation method,
by which we obtain another result for the scaling dimension from same data.
If the two results are similar,
our extrapolation method (or fit function) to the continuum limit works well.

The new method is based on the excision of a small region around the contact point
of the integrand~$\langle A(x)A(0) \rangle$
in~$\ln\chi(L)$ in~Eq.~\eqref{eq:4.1}~\cite{Kawai:2010yj}.
The modified scalar susceptibility~$\Tilde\chi$ is defined by
\begin{align}
 \Tilde\chi(L)
 = \frac{1}{a^2} \int_{|x|\geq \lambda^{-1}} d^2 x\,
 \left\langle A(x) A^{*}(0) \right\rangle .
 \label{eq:4.11}
\end{align}
The coupling~$\lambda$ is the unique mass scale
in the WZ model with the superpotential in~Eq.~\eqref{eq:2.1},
and the correlations at short lengths~$\sim\lambda^{-1}$ would not affect
the scaling in~Eq.~\eqref{eq:4.3} of~$\chi(L)$ in low-energy regions.
Note that the shape of the excised space is slightly different from
those in~Refs.~\cite{Kawai:2010yj,Kamata:2011fr},
but the susceptibility should not be sensitive to such UV details
\textit{in the continuum limit};
if the grid size~$L/a$ is not sufficiently large (i.e. $L/a$ is finite),
we suffer from sensitivity to the excised space size;
this is the problem that the susceptibility in~Ref.~\cite{Kamata:2011fr}
is quite sensitive to the UV ambiguity.
In terms of the Fourier modes of~$A$, we have
\begin{align}
 \Tilde\chi(L)
 &= \frac{1}{a^2L^2} \left\langle |A(p=0)|^2 \right\rangle
 - \frac{1}{a^2L^4} \sum_p \frac{2\pi\lambda^{-1}}{|p|}
 J_1 (\lambda^{-1}|p|) \left\langle |A(p)|^2 \right\rangle ,
 \label{eq:4.12}
\end{align}
where $|p|=\sqrt{p^2}$ and~$J_1$ is the Bessel function of the first kind.

The parameter tuning above indicates
that the dimensionless coupling~$a\lambda$ becomes large as $L/a\to\infty$,
while~$\ln\chi(L)$ is kept fixed.
That is, in the small-$a$ limit,
the volume of the excised space becomes smaller and smaller;
we must have completely the same result of the scaling dimension
as in the method of~Eq.~\eqref{eq:4.5}, at least analytically.
In numerical simulations, however, it is not known a priori
what function we should apply to take the continuum limit.
Thus, attempting to extrapolate results of~$\ln\Tilde\chi(L)$
and to determine the fit function,
one can justify the numerical determination of the scaling dimension from~$\Sigma$.
In the same way as~$\ln\chi(L)$,
we define the new scaling function~$\Tilde\Sigma$ by
\begin{align}
 \Tilde\Sigma(s, u, a/L) = \ln\Tilde\chi(s L) .
 \label{eq:4.13}
\end{align}
Here, $u$ is given by the fixed number $\ln\chi(L)$,
which is identical to the value of~$\ln\Tilde\chi(L)$ in the continuum limit,
that is, $\lambda^{-1}\to0$.
Similarly, one can measure the scaling dimension by~Eq.~\eqref{eq:4.5}
with~$\Tilde\Sigma$ and another to-be-determined fit function.

\begin{figure}[t]
 \begin{center}
  \includegraphics[width=0.8\columnwidth]{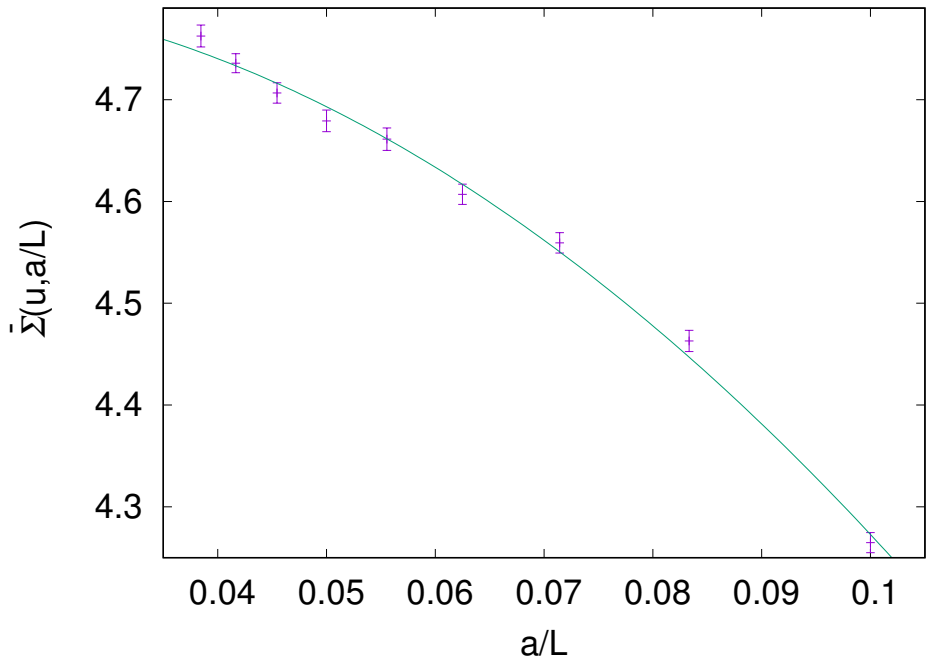}
  \caption{$\Tilde\Sigma(u,a/L)$-$(a/L)$ plot with $u=3.9175$.
  The fitting curve of~Eq.~\eqref{eq:4.14} is also depicted.}
  \label{fig:chi2}
 \end{center}
\end{figure}

From the $\Tilde\Sigma(u,a/L)$-$(a/L)$ plot in~Fig.~\ref{fig:chi2}
we obtain the fitted quadratic curves
\begin{align}
 \Tilde\Sigma(3.9175,a/L)
 &= -0.091(14) \times \left(\frac{26a}{L}\right)^2 + 0.031(52) \times \frac{26a}{L}
 + 4.8062(425)
 \label{eq:4.14}
\end{align}
with~$\chi^2/\text{d.o.f.} = 1.600$, or
\begin{align}
 \Tilde\Sigma(3.9175,a/L)
 &= -0.0823(19) \times \left(\frac{26a}{L}\right)^2 + 4.8317(62)
 \label{eq:4.15}
\end{align}
with~$\chi^2/\text{d.o.f.} = 1.423$.
These fitting results give the scaling dimension as
\begin{align}
 1-h-\Bar{h} &= 0.641(31) ,&
 1-h-\Bar{h} &= 0.6594(45) ,
 \label{eq:4.16}
\end{align}
respectively.
These two results are consistent with our previous result in~Eq.~\eqref{eq:4.10}.
We have obtained the precise and reliable result in~Eq.~\eqref{eq:4.10}
through the finite-size scaling with the continuum-limit extrapolation.

\section{Conclusion}
In this paper we numerically studied the IR behavior
of the~$2$D~$\mathcal{N}=2$ WZ model with the cubic superpotential,
which is believed to provide the Landau--Ginzburg description
of the~$A_2$ minimal model of the~$2$D~$\mathcal{N}=2$ SCFT.
To take the continuum and infinite-volume limits,
we developed a systematic extrapolation method for the scalar susceptibility~$\chi(L)$;
this method is applicable to various non-perturbative formulations of the model.
Then, from the numerical simulation of~$\chi(L)$
on the basis of the SUSY-invariant formulation with a momentum cutoff,
we performed the precision measurement of the scaling dimension
through the finite-size scaling analysis.
The result of the scaling dimension in~Eq.~\eqref{eq:4.10} is rather consistent
with the conjectured WZ/SCFT correspondence.

As shown in~Table~\ref{tab:h_finite} and~Fig.~\ref{fig:chi},
we observed a significant difference
between our net result and the ones at any finite~$L/a$.
The scalar susceptibility takes a slow approach to the~$a/L=0$ limit,
at least in the present formulation.
By using our extrapolation analysis,
we can get down to the target SUSY continuum theory with the infinite volume;
from a numerical simulation based on the formulation by Kadoh and Suzuki,
we obtained the limiting value for the simplest~$A_2$ theory.
This result not only has a smaller margin of error in the numerical value,
but also would be much more reliable than those of preceding studies,
which were computed at finite~$L/a$;
it shows a coherence picture being quite consistent with the theoretical conjecture.

Our result seems to support the restoration of the locality in the continuum limit.
The UV ambiguity in~$\chi(L)$ with finite~$L/a$,
that is, the sensitivity to the excised space size~$\sim\lambda^{-1}$
around the contact point, has disappeared because~$\lambda^{-1}\to0$ in the limit.
We indeed found that
the results in~Eq.~\eqref{eq:4.16} based on the excision prescription
are consistent with~Eq.~\eqref{eq:4.10} without the excision.
Also, in addition to the earlier numerical simulations based on the present formulation,
it would be exemplified by
good agreement between~Eq.~\eqref{eq:4.10} and the expected value
that the momentum-cutoff regularization in the~$2$D theory works quite well.
However, the theoretical background of our computational approach is still not clear,
so we should observe the locality restoration more carefully;
this is a future problem.

A related issue is the continuum-limit analysis of the central charge.
Such an analysis will be useful to study general SCFTs.
It is important to confirm further the theoretical validity of the formulation,
in order to investigate superstring theory via the LG/Calabi--Yau correspondence.

\section*{Acknowledgments}
We would like to thank Sinya Aoki, Daisuke Kadoh, Yoshio Kikukawa, Taichiro Kugo,
Katsumasa Nakayama, and Hiroshi Suzuki for helpful discussions and comments.
The numerical computations were partially carried out by the supercomputer system ITO
of the Research Institute for Information Technology (RIIT) at Kyushu University.
This work was supported by JSPS KAKENHI Grant Number JP18J20935.

\end{document}